# Transformation of deep-water methane bubbles into hydrate

Alexander V. Egorov[1], Robert I. Nigmatulin[1], Aleksey N. Rozhkov[1,2]

[1]*P.P. Shirshov Institute of Oceanology, Russian Academy of Sciences, 36 Nahimovski prospect, Moscow 117997, Russia*

[2]*A. Ishlinsky Institute for Problems in Mechanics, Russian Academy of Sciences, 101(k.1) Prospect Vernadskogo, Moscow 119526, Russia*

E-mail: avegorov@ocean.ru

**ABSTRACT**

The paper is dedicated to the mechanics of the methane bubbles in the gas hydrate stability zone of the basin. Transformation of deep-water methane bubbles into solid hydrate was investigated in Lake Baikal *in situ*. Released from the bottom methane bubbles were caught by different traps with transparent walls. It was observed that when bubbles entered into internal space of the trap, the bubbles could be transformed into two different solid hydrate structures depending on ambient conditions. The first structure is hydrate granular matter consisted of solid fragments with sizes of order of 1 mm. The second structure is high porous solid foam consisted of solid bubbles with sizes of order of 5 mm. The formed granular matter did not change during trap lifting up to top border of gas hydrate stability zone, whereas free methane intensively released from solid foam sample during it lifting. It was concluded that the decrease of the depth of bubble sampling and the decrease of the bubble flux rate assist to formation of hydrate granular matter, whereas increase of the depth of bubble sampling and growth of bubble flux rate assist to conversation of bubbles into high porous solid hydrate foam.

# Introduction

The present work was undertaken to study the behavior of deep-water methane bubbles at the water reservoir depths at which the methane hydrates can be formed and can be thermodynamically stable. This is so-called hydrate stability zone - HSZ. In HSZ a phase transition "gas - hydrate" is possible. In this process each molecule of the gas turned out to be in a "cage" of few water molecules. In the case of methane the average number of water molecules is about six. So way formed construction is solid and inflammable in the air.

The purpose of the work is to answer the question, what kinds of solid structures of methane hydrates can be formed from deep water methane bubbles as a result of the gas-hydrate transition. The types of the formed structures can define the rate of methane hydrate dissolution in the water and therefore can influence on global climate change by means of control of methane fluxes from bottom to surface as it discussed by Judd *et al*. (2002) and MacDonald *et al*. (2002). Another object of the influence is the safety of the underwater buildings which can be failed by means of undesirable phase transformations (Hagerty & Ramseur 2010). Knowledge about the mechanics of phase transformations of deep methane bubbles is crucial for optimum and safe choice of a technology during underwater work in the deep waters of the World Ocean.

Experiments were carried out during the expedition of Russian Academy of Sciences "MIRI NA BAIKALE 2008-2010." In the Lake Baikal the water on the bottom has low temperature. Besides here water is subjected to high hydrostatic pressure due to the large depth of the lake. There are seeps of methane in the form of bubbles from the sediment into the hydrosphere in some deep places. Thus, there are conditions in Lake Baikal for transfer of the gas into the solid hydrate state. All these features were used to study the possible transformation of the methane bubbles into solid hydrate structures. Experiments were carried out with the bubbles in the Lake Baikal with deep-water manned submersible (MS) "MIR" - Fig. 1.

Preliminary results were published by Egorov *et al*. (2010), Egorov *et al*. (2011) and Egorov *et al*. (2012). Current paper presents only new photographs obtained from video records in other moments of time. Additional experimental observations and experimental data were analyzed. As a result corrected numerical estimates were obtained. Experiment with non-natural methane is reported. Modeling was modified and significant alteration of previous results was made. A generalization and physical explanations of the observed effects are suggested in the present work.

In particular for the first time the statement about two controlling factors of gas-hydrate transformation was supposed and discussed. Comments regarding bubble behavior at extra high depth were re-considered.

To collect the gas bubbles several versions of traps were used - Fig. 2 (Egorov *et al.* 2010; Egorov *et al.* 2011; Egorov *et al.* 2012). Trap "Funnel" is an inverted glass, into which the funnel is inserted. The sampled gas or solid hydrate formation remained in the trap, even when it is turning. Trap "Grid" is an inverted glass, divided into sections by one grid (version "Grid-1") or by two grids (version "Grid-2"). Appointment of grid is to determine has the bubble hydrate envelope or not. Hydrate envelope prevents the pass of gas through the grid and as well as the coalescence of bubbles during their contact. The side walls of the trap "Grid" had small holes (ϕ4 mm) for the outflow of excess water in the course of the gas entering into the trap.

By means of mechanical arm of MS "MIR" the trap was located near the bottom, so that the stream of bubbles from the bottom fed into a trap.

## Results

Submersion of MS "MIR" to experiment with deep methane bubbles were conducted in Lake Baikal at three areas: "Goloustnoye" (51°58.42 N, 105°20.99 E, 3 July, 2010), "Gorevoy Utes "(53°18.35 N, 108°23.57 E, 19 August, 2009) and "St. Petersburg" (52°52.97 N, 107°09.99 E, 30 July, 9 and 11 August, 2010) - Fig. 1. The main difference between areas is their different depths. It was 405, 860 and 1400 m in areas, respectively. The water temperature on the bottom varied in the range +3.5÷4.0 °C depending on the depth. The intense release of methane bubbles from the bottom was observed in all three areas at the dates indicated. The upper boundary of HSZ in Lake Baikal is determined at a depth $z_s$~380 m (Egorov *et al.* 2011).

Area "Goloustnoye", is located near the village of "Bol'shoe Goloustnoye", 50 km north-east of the source of the Angara river. Date of dive was July 3, 2010. The water temperature at a depth of 405 m (bottom) was about 4 °C. Therefore, temperature and pressure conditions at the bottom satisfy to the criterion of stability of methane hydrates with minimal odds. There was a strong natural release of methane bubbles at the area.

Collection of bubbles by means of a trap "Grid-1" with 1 mm mesh grid did not reveal differences from that of conventional gas bubbles (Fig. 3). Bubbles delayed by grid and then coalesced with each other. As a result a flat single bubble is formed under the grid. As the result of the accumulation of a sufficient amount of gas under the grid and/or shaking of the trap the gas, overcoming the Laplace capillary pressure, flowed through the grid into upper section of the trap. Here gas once again formed a single flat bubble. This behavior of trapped bubbles demonstrates the evidence of the absence of any film (hydrate, surfactant, etc.) on the surface of rising bubbles.

Observations of the collection of bubbles with trap "Funnel" at depth 405 m also did not reveal any deviations from usual behavior of conventional gas bubbles in the water. The trap is gradually filled by gas, displacing the water. Any traces of solid hydrates were not found.

However, as the result the MS motion along bottom relief up to more deep depth of 550 m, a thin layer of hydrate was formed on the inner surface of the trap, which was in the gaseous phase (Fig. 4). If sometimes the trap was overturned the hydrate was formed also on the wet surface of the funnel during the time when the surface of funnel was in the gas phase.

Thus, the observations at the area "Goloustnoye" demonstrated impossibility or significant slowing down of gas hydrate formation from free methane in HSZ if temperature and pressure conditions are close enough to critical ones. The reasons for such behavior are not clear so far.

Note also that the numerous attempts to form hydrate in the bottle with non-natural (industrial) methane by means of deep immersion of this bottle in the lake up to a depth of 1400 m were undertaken. Only in few cases attempts led to the appearance of small hydrate fragments in the bottle. In these experiments before submersion the bottle (~200 ml) and attached gas reservoir (~1500 ml) were filled with gas. The wide hole in the bottom of the gas reservoir connects the gas with the hydrosphere. During the submersion the hydrostatic pressure increases and therefore the gas is compressed in many (~140) times. The water fills the reservoir and bottle through the hole in the bottom of the reservoir. Figure 5 demonstrates unique example when after a 5-hour stay at the depth of 1400 m methane hydrate film was formed at the water-gas border. Hydrate formation can be proved by the absence of any disturbance of the border under the shaking of the bottle.

Area "Gorevoy Utes" is located near the Cape "Gorevoy Utes" at coordinates 53°18.35 N, 108°23.57 E. Experiments were conducted 19 August, 2009. The lake depth here is 864 m, water temperature at the bottom is +3.5 °C, .i.e., pressure and temperature conditions at the bottom obviously meet the stability of methane hydrates.

Experiments with trap "Grid 1" (one grid in the middle of the trap, size of mesh cell is of order of 1 mm) showed that bubbles of methane are delayed by grid - Fig. 6. Coalescence of bubbles did not occur as it occurred in the previous test area "Goloustnoye". Absence of coalescence indicates that the rising bubble is covered by solid hydrate envelope. Bubbles gradually transformed into hydrate granular matter - hydrate powder. Hydrate granular matter has positive buoyancy and its particles are arranged under the grid. The shaking of the trap changed granular matter distribution in the trap. Some part of the granular material sifted through a 1 mm mesh and located in the upper section of the trap. The other part (consisted of larger particles >1 mm), remained in the lower section of the trap under the grid.

When the trap was oscillated by mechanical arm of MS the hydrate granular matter freely interspersed in the trap sections every time occurring in the tops of sections - Fig. 7.

In experiments with trap "Funnel" it was found that trapped bubble remained in its original form for a few minutes, losing transparency and becoming opaque. The formation of solid hydrate envelope of the bubble occurred. A few minutes later hardened bubble collapsed into a number of small hydrate fragments. Sequential breakup of bubbles formed so way a white granular matter - powder of hydrate particles of different sizes (Fig. 8). Having positive buoyancy, the forming granular matter accumulated at the top of the trap.

Under trap oscillations the granular matter freely interspersed from one end to another end of the traps - Fig. 9.

During few hours stay at depth of 860 m and the subsequent MS ascent up to a depth of 387 m not any changes in formed hydrate granular matter were observed. Not any escape of free gas from hydrate granular matter in the trap was observed too. This indicates the absence of the gas phase inside the hydrate structure and probably complete transformation of the original methane bubble into solid hydrate. The subsequent MS ascent above the depth of 387 m, i.e. above HSZ upper boundary, caused the hydrate granular matter decomposition and it transformation into free methane gas that filled the trap - Fig. 10.

Area "St. Petersburg" is a deep mud volcano "St. Petersburg." Coordinates of volcano are 53°52.97 N, 107°09.99 E. Depth here is equal 1400±10 m. Studies of methane bubbles at the area conducted in the summer of 2010 in a series of submerges.

When bubbles entered into the trap "Grid" or "Funnel", the bubbles did not coalesce with each other (as at a depth of 405 m), not collapsed (as at the depths of 860 m), but bubbles formed a solid hydrate foam. Absence of coalescence and correspondent foam formation clearly indicates the existence of the hydrate envelope on the surface of the original bubble.

One can see a distinct difference between hydrate foam and hydrate granular matter, observed at the area "Gorevoy Utes." Foam structure remained stable for the duration of the submersion. Initially, the foam in the trap "Funnel" looked like quite elastic foam - Fig. 11. Under trap shaking or overturning the foam slightly deformed, preserving, however, the same configuration. Such foam behavior resembled the behavior of conventional foam in the process of it motion through channel of variable cross section (Bazilevsky & Rozhkov 2012). It was possible to achieve the foam detachment from trap wall by mean of strong trap shaking. But foam separation occurred only in the form of one or a few macro-pieces of foam. If pieces of foam are left alone, then over time they again attached to the walls of the trap and only a strong shaking could detach foam from the walls of the trap. Foam was relatively transparent. Well-defined bubbles were seen inside the foam. The sizes of these bubbles in the foam corresponded to ones of original bubbles before their entering into the trap ~ 3-7 mm. Besides large volumes of free gas were seen in the trap. The sizes of these gas volumes exceeded the sizes of initial bubbles. It is likely that these volumes have been formed during periods of very strong fluxes of bubbles into the trap. High perturbations of the bubble stream and convergent character of bubbler motion in the funnel assisted to coalescence of part of bubbles in spite of the resistance of the hydrate layers on the surfaces of the bubbles.

Later hydrate foam in the trap became harder, losing elasticity - Fig. 11. Solid cellular structure occupied almost all space of the foam.

After entering of bubbles into the trap "Grid-2" the bubbles delayed by the lower grid with the cell of 6 mm as well as by the upper grid with a cell of 1 mm. However, some time later due to fluctuations of the bubble stream and/or some trap shaking the part of bubbles overcame grids, moving upward in the trap. In the end, a grid of 6 mm cell was practically free from bubbles, and most of the bubbles were concentrated in the middle and upper parts of the trap (Fig. 12).

The trap content was monitored during the MS ascent. A great difference between the behavior of hydrate foam (formed at a depth of 1400 m) and the behavior of hydrate powder (formed at a depth of 860 m) in the process of ascent was observed:

Hydrate powder remained unchanged during the ascent up to a depth of 380 m (top HSZ boundary) as it discussed above.

The hydrate foam released a free gas just after start of the MS lifting. The volume of released free gas increased with decreasing of the depth, i.e. with the decreasing of the hydrostatic pressure.

The growth of the free gas volume is indicated by the movement of the meniscus down - Fig. 13. The pictures also show a slight increase of the size of solid foam sample.

Obviously, in the upper section the meniscus can only move down. Water from the upper section is displaced into the surrounding hydrosphere through a plastic tube (Figs. 12, 13).

In the middle section expansion of the gas proceeds mainly by moving of the meniscus down, although a small amount of gas in the form of a few bubbles released from the foam volume upward, penetrating through the grid of 1 mm cell. Preferential flow of gas down indicates a lower permeability of the upper layers of foam compared to the lower ones. The upper layers of foam were formed earlier, so they had more time to build a more hard hydrate structure.

Video analysis of displacement of the meniscus in the upper and middle sections of the trap "Grid-2" during MS ascent allows recovering the variation of the volume occupied by the gas and the solid hydrate phase in the upper and middle sections - Fig. 14. Plot also presents similar data obtained with the use of traps "Thermo" (Egorov *et al.* 2012). Notations are used: $V=v/v_0$, where $v$ is the total volume occupied by the gas and hydrate foam, $v_0$ is the volume occupied by the hydrate foam at the beginning of the ascent of MS, $Z=z/z_0$, $z_0=1400$ m is the initial depth, $z$ is the current depth.

To analyze the data a simple model of the phenomenon is suggested. It is assumed that during MS ascent the mass of the gas in the section (upper or middle) of the trap is varied as

$$m=m_0+m_n(1-z/z_0) \qquad (1)$$

due to consumption of gas for additional hydrate formation ($m_n<0$) or due to release of gas as a result of the decomposition of the original hydrate ($m_n>0$). As a first approximation a linear dependence is proposed, in which $m_0$ and $z_0$ are the mass of gas and the depth at the start of the MS ascent.

Gas and solid hydrate occupied in the trap section the volume $v$, which is the sum of the volume of free gas $v_g$ and volume of the solid hydrate phase $v_{gh}$: $v=v_g+v_{gh}$. At the moment of the ascent start $v=v_0$, $v_g=\phi_0 v_0$, $v_{gh}=(1-\phi_0)v_0$, where $\phi_0$ is the porosity of the solid foam.

Variation of the volume of gas in the trap section during the ascent is described by the Clapeyron-Mendeleev equation of state:

$$v_g=(m/\mu)R_g T/p \qquad (2)$$

where $\mu$ is the molar mass of methane, $R_g$ is the universal gas constant, $T$ is the thermodynamic temperature, $p$ is the hydrostatic pressure. Gas volume $v_g$ changes due to changes of the mass $m$ and the hydrostatic pressure $p$.

The volume of the solid hydrate in the trap section during the ascent varies only due to hydrate decomposition ($m_n>0$) or due to the formation of a new portion of a hydrate ($m_n<0$):

$$v_{gh}=(1-\phi_0)v_0-m_n(1-z/z_0)/(\rho_{gh}\kappa) \tag{3}$$

where $\rho_{gh}$ is the density of methane hydrate, which is independent on pressure, $\kappa=16/(16+6\times18)=0.129$ is the mass fraction of methane in gas hydrate, in which one molecule of methane (molecular mass 16) is connected (in average) with six molecules of the water (molecular mass 18).

The solution of the algebraic system of equations (1) - (3) is:

$$V=\phi_0(p_0/p)+(1-\phi_0)+M_n(1-Z)(\phi_0(p_0/p)-1/A) \tag{4}$$

where $V \equiv v/v_0$, $p_0=p_a+\rho g z_0$ is the hydrostatic pressure at the start of the ascent, $p=p_a+\rho g z$ is the current hydrostatic pressure, $p_a$ is the atmospheric pressure, $\rho$ is the density of water, $g$ is the acceleration of gravity, $p_0/p=(1+P_a)/(Z+P_a)$, $P_a=p_a/\rho g z_0$, $M_n=m_n/m_0$, $Z=z/z_0$, $A=\rho_{gh}\kappa R_g T/\mu p_0$. Taking $R_g$=8.31 J/(mol K), $\rho_{gh}$=900 kg/m$^3$, $\rho$=1000 kg/m$^3$, $T$=273 K, $\kappa$=0.129, $\mu$=0.016 kg/mol, $p_a$=10$^5$ Pa, $g$=9.81 m/s$^2$, $z_0$=1400 m, we find that for a given depth of 1400 m $P_a$=0.0073, $A$=1.19.

Obviously, equation (4) is Boyle-Mariotte law with the addition, taking into account the release/consumption of gas in the trap section. Formula (4) was used for fitting of the experimental data in Fig. 14. Variables $\phi_0$ and $M_n$ were used as fitting parameters. Fitting procedure was applied to the experimental points lie within HSZ $z<z_s$=380 м ($Z<Z_s$=0.2714). The correspondent data are marked in the plot by points within the symbols. Fitting was carried out by means of least squares methods using graphics software package Origin 6.1. The calculations were performed in the mode "No weighting". Not any restrictions were imposed on the parameters $\phi_0$ and $M_n$. The following results of fitting were obtained:

$$\phi_0=1.08404\pm0.18128, M_n=0.19296\pm0.29101 \tag{5}$$

Approximation (4) with fitting parameters (5) is in good agreement with the experimental data - Fig. 14. Agreement continues be valid beyond HSZ.

Figure 14 also shows of the dimensionless form of the dependence (1)

$$M(\equiv m/m_0)=1+M_n(1-Z)$$

with parameters (5). The dotted lines indicate the confidence interval of possible variation of the mass of gas in the trap $M$.

If we admit a priory the impossibility of hydrate decomposition into water and free gas in the HSZ $M_n<0$ (although there is an alternative point of view (Khlystov *et al*. 2010)) and impose the formal restriction $\phi_0<1$, then the fitting with equation (4) defines the fitting parameters as:

$$\phi_0=1^{+0}_{-0.07927}, M_n=0^{+0}_{-0.00835} \qquad (6)$$

Result (6) does not contradict to the previous solution (5). Consequently, the results of both fittings of the experimental data show very high foam porosity $\phi_0\cong1$ and the conservation of mass of gas in the trap $M\cong1$ during its motion from the bottom to the surface.

Thus a physical model of the phenomenon on the basis of Boyle-Mariotte law provides a satisfactory description of the experimental observations - Fig. 14. So direct measurements revealed that during the ascent the free gas expands according to Boyle-Mariotte law, in which the initial gas volume is equal to the total volume of foam formed at the initial depth.

## Discussion

The main question that arises in the analysis of the presented observations is: Why the methane bubbles at the area "Gorevoy Utes" formed a single-phase hydrate granular matter, but at the area "Saint Petersburg" bubbles formed highly porous hydrate foam? At the moment, our working hypothesis is a statement that two factors control the transformation of methane bubbles in the traps in HSZ.

The first factor is a reduction of pressure inside the bubble due to the consumption of gas for the formation of the hydration envelope. Reduction of pressure in the bubble forms the inner stresses in the envelope that can destroy envelope. The second factor is the rate of gas bubbles flux in the trap, which defines how long the incoming bubble is in the aqueous phase, before it will be surrounded by other bubbles. Thus these factors control the mechanisms of methane bubble transformation into solid hydrate.

If a single bubble is a long time in an aqueous environment, then the development of internal stresses in the envelope can destroy the bubble envelope, i.e. the first factor dominates. On the other hand if the trapped bubble is quickly surrounded by other bubbles then bubble loses the contact with the water, the formation of the hydration envelope stops and a bubble becomes stable element of formed foam. So, what will happen with methane bubbles in the trap depends on the competition of these two factors.

## First controlling factor of bubble transformation

Kinetics of hydrate envelope formation and the corresponding gas consumption rate are described by the equations of mass conservation and Clapeyron-Mendeleev state equation (Fig. 15)

$$dm/dr = 4\pi r \rho_{gh} \kappa \qquad (7)$$

$$p_1(4/3\pi r^3) = (m/\mu) R_g T \qquad (8)$$

where $m$ is the mass of free gas in the bubble, $\kappa=0.129$ is the mass fraction of methane in the hydrate, $\rho_{gh}$ is the density of methane hydrate, $\mu$ is the molar mass of methane, $R_g$ is the universal gas constant, $T$ is the thermodynamic temperature, $p_1$ is the pressure inside the bubble, $r$ is the internal radius of the hydrate envelope of the bubble. It is assumed that the water can be filtered into the bubble through the cracks and micro-channels in the hydrate envelope (Ribeiro & Lage 2008). As soon the hydrate formation decreases the pressure in the bubble, the water filtration is supported by formed pressure drop. Hydrate is deposed on the internal surface of the bubble.

Integration of equations (7) and (8) with the initial conditions $p_1=p_0$, $r=r_0$ ($p_0$ is the hydrostatic pressure at a given depth, $r_0$ is the initial radius of the bubble) determines the variation of the pressure inside the bubble with growth of hydrate envelope inside the bubble

$$P_1 = A + (1-A)/R^3 \qquad (9)$$

where $P_1=p_1/p_0$, $A=\rho_{gh}\kappa R_g T/\mu p_o$, $R=r/r_0$.

Equation (9) has a physical meaning only when $P_1 > 0$. Thus there is a limiting radius $R_f=(1-A^{-1})^{1/3}$, at which hydration stops, since all methane has already been consumed.

At the depths of HSZ contribution of atmospheric pressure in the hydrostatic pressure is relatively small ($P_a \ll 1$), so $p_0 \cong \rho g z_0$. In this case the value of $A$ can be represented as a ratio of two depths $A = z_*/z_0$, one of which is a transitional depth $z_* = \rho_{gh} \kappa R_g T / \mu \rho g$ and other is the actual depth $z_0$. Assuming $\rho_{gh}$=900 kg/m³, $\kappa$=0.129, $\rho$=1000 kg/m³, $R_g$=8.31 J/(mol K), $T$=273 K, $\mu$=0.016 kg/mol, $g$=9.81 m/s², we obtain that $z_* \approx$1678 m. At this depth the methane density $\rho_m$ ($\rho_m = p_*/(R_g T/\mu)$, $p_* = \rho g z_*$) differs from one of methane hydrate $\rho_{gh}$ exactly in $\kappa$ times, i.e. $\rho_m = \kappa \rho_{gh}$. In the course of the reaction of methane with water at the depth $z_* \approx$1678 m the volume of methane reacted is exactly the same as the volume of hydrate formed. For example, the formation of a layer of hydrate d$r$ (Fig. 15) will consume only the methane gas that occupied the layer d$r$ before. Therefore, the growth of hydrate does not affect the state of the gas in the central part of the bubble as long as the envelope will not reach the centre of bubble.

When $z_0 < z_*$ ($A < 1$) the growth of the hydrate envelope causes a pressure reduction in the bubble, as described by equation (9). Pressure reduction is more noticeable if the actual depths $z_0$ is close to the upper boundary of HSZ $z_s$ (parameter $A = z_*/z_0$ is small). Pressure decreases with growth of hydrate envelope when variable $R$ increases.

In the framework of the model, the growth of the envelope is possible until all gas is consumed $P_1$=0. According to (9), gas is over when the radius of the envelope reaches

$R_f = (1 - z_0/z_*)^{1/3}$

For depths $z_0$=405, 860 and 1400 m limiting radius is $R_f$=0.9120, 0.7870 and 0.5492, accordingly. The greater the depth, the thicker envelope can be formed. Indeed, at a greater depth the bubble contains more gas, which can be spent for building of thicker envelope.

On the other hand the model predictions for real methane are limited by thermodynamic instability of hydrate at lower pressure. Indeed, as soon as the pressure inside the bubble $p_1$ will fall up to a level of critical pressure $p_s$, then further growth of the hydrate envelope will stop. Critical pressure $p_s$ is a minimal pressure at which hydrate is stable at a given temperature. At pressure $p_s$ any additionally formed hydrate decreases pressure in the bubble below level $p_s$ and therefore the addition decompose immediately. For the temperature of the waters of Lake Baikal $T$=3.5 °C the critical phase transition pressure $p_s$, equal $p_s = p_a + \rho g z_s$=3.8278 MPa, where $z_s$=380 m is the upper boundary of HSZ (Egorov et al. 2011). Substitution $p_1 = p_s$ ($= \rho g z_s$, $p_a \ll \rho g z_s$) in (9) determines the maximal envelope radius $R_s$, at which further growth of hydrate in the bubble stops because of the thermodynamic instability:

$$R_s = ((1-z_0/z^*)/(1-z_s/z_*))^{1/3}$$

i.e. $R_s = R_f/(1-z_s/z_*)^{1/3} = R_f/0.9180$ for conditions of Lake Baikal. For depths $z_0 = 405$, 860 and 1400 m radius of the thermodynamic instability $R_s = 0.9935$, 0.8573 and 0.5983.

Pressure difference $p_0-p_1$ between the environment and the gas phase inside the bubble (equation (9)) generates inner stresses in the envelope, the values of which are described by the exact solution of Lamé (Timoshenko & Goodier 1970)

$$\sigma_t - \sigma_r = 3/2 \rho g z_0 (1 - z_*/z_0) \qquad (10)$$

at the internal surface of the envelope, and

$$\sigma_t - \sigma_r = 3/2 \rho g z_0 (1 - z_*/z_0)/R^3 \qquad (11)$$

at the external surface of the envelope. In formulas $\sigma_r$ is the radial stress, $\sigma_t$ is the circumferential (tangential) stress (Fig. 15). Factor $\rho g z_0(1-z_*/z_0)$, obviously is the difference between hydrostatic pressures at the current depth $z_0$ and at the transitional depth $z_*$.

The difference between normal stresses is presented here as magnitude which defines the destruction threshold. Indeed, as a first approximation namely the level of normal stresses difference provides the transition to the destruction under compression. This statement is based, for example, on the theory of the maximal shear stress of Tresca (Sedov 1997), which is one of the most recognized destruction theories in geo-mechanics.

Figure 16 shows the inner stress in the envelope $\sigma_t - \sigma_r$ as function of the depth $z_0$ and thickness of the envelope (parameter $R$). The calculations are performed according to (10), (11). The envelope radius $R_f$ and $R_s$ are demonstrated.

The actual depth $z_0$ is closer to the HSZ boundary $z_s = 380$ m, the stress level $|\sigma_t - \sigma_r|$ is greater. The stresses at the internal and at the external surfaces of the envelope coincide in the moment of the hydrate envelope birth. The subsequent growth of the envelope does not change the stress at the external surface of the envelope, but the stress at the internal surface increases. The depth is shallower, the stress is growing faster. However, at all depths the same maximum value of the stress can be achieved. Depending on the mechanism of the process end (consumption of all gas or thermodynamic instability) maximal level of the stress is equal $|\sigma_t - \sigma_r| \sim 25$ MPa, or $|\sigma_t - \sigma_r| \sim 19$ MPa. Indeed, the substitution of $R_f = (1-z_0/z_*)^{1/3}$ in equation (11) leads to $\sigma_t - \sigma_r = -3/2 \rho g z_* = -24.69$ MPa, and the substitution $R_s = ((1-z_0/z_*)/(1-z_s/z_*))^{1/3}$ leads to the stress $\sigma_t - \sigma_r = -3/2 \rho g (z_* - z_s) = -19.10$ MPa.

Thus, the calculations show that as a consequence of the transition from a depth of 1400 m to a depth of 860 m the level of stresses in the envelope significantly increases. If the stress exceeds the compression strength of the hydrate material, the material breaks down into separate fragments. Remaining gas forms a bubble again, then collapsing again by the same mechanism.

According to the presented calculations, much more favorable conditions for the destruction of the hydrate envelope are available at a depth of 860 m than at depths of 1400 m. Stress magnitude at a depth of 860 m (see data in Fig. 16) substantially higher than, for example, the strength of the ice under the compression. It can be proposed that strength properties of hydrate are close enough to the strength properties of the ice because both materials are similar enough. Strength of the ice under the compression is estimated to be ~10 MPa. Therefore, a hydration envelope at a depth of 860 m can be subjected to the brittle fracture. Apparently at a depth of 1400 m the stress level is not sufficient to fracture hydrate; here the process of methane transformation into hydrate follows to a different scenario. As for the depth of 405 m, the hydrate formation is not observed, so there is no data about possible hydrate forms at this depth.

Mechanics of formation and possible destruction of the hydration envelope of the bubble at depths below the transitional depth $z_*=1678$ m is not clear. In the framework of the model the pressure increases in the bubble at these depths and prevents to filter water into the bubble through the envelope. On the other hand, the alternative assumption about the growth of hydrate on the external surface of the envelope leads to the conclusion that the pressure decreases inside the bubble and therefore cannot support the methane filtration through envelope. Perhaps, only experimental observations may clarify the physical picture of the phenomenon.

## Second controlling factor of bubble transformation

The second factor that influences on the mechanism of the methane bubbles transformation into hydrate is the rate of bubbles flux (Leifer & Culling 2010) into the trap. At flux of low intensity the bubble remains alone in the trap long time. Bubble has enough time to be covered by the hydrate envelope of finite thickness. It can collapse according to the above scheme if the depth allows formation of inner stress, which magnitude exceeds the strength of the hydrate.

If the depth is not sufficient for bubble collapse then another possibility for a single bubble arises. Bubble can wait for the arrival of the next few bubbles and take part in the formation of the so-called spherical foam (Adamson 1982). Gas bubbles are in water media and separated by relatively thick hydrate envelops - Fig. 17. The level of internal pressure in the bubbles is insufficient to break the envelope.

The gas release from spherical foam during MS ascent does not follow to Boyle-Mariotte law, because the porosity of spherical foam is not small. Thus one can conclude that the spherical solid hydrate foam was not formed in experiments at area "St. Petersburg". At area "St. Petersburg" methane bubbles transformed into highly porous hydrate foam (polyhedral foam, see below) with a minimal content of the solid phase and zero content of the water.

Highly porous hydrate foam is an example of polyhedral foam (Adamson 1982). A foam cells are separated by thin hydrate walls (Fig. 17). Each cell looks like polyhedra. It is assumed that the polyhedral solid hydrate foam is formed due to the high intensity of the bubbles flux into the trap. Coming bubbles collide domestic bubbles and squeeze them under the force of Archimedes. At high intensities, bubbles do not have time enough to form a thick hydration envelope (Rehder *et al* 2009). So hydrate envelope can be easily deformed when bubbles contact with each other. Bending deformations of elastic envelopes displace water from foam space, as in the case of conventional foam drainage (Jun *et al*. 2012). Conventional polyhedral liquid foam is provided by disjoining pressure in thin lamellae of foam (Kornev et al 1999), the polyhedral solid hydrate foam is provided by a thin hydrate layer on the boundary of the gas phase. A thin hydrate layer plays the role of disjoining pressure, preventing bubble coalescence.

## Conclusions

Summarizing the above comments regarding mechanisms of the transformation of methane bubbles in HSZ, the following conclusions can be suggested. Depending on ambient conditions the methane bubbles can be transformed into a granular matter, or form solid hydrate foam, or remain conventional gas bubbles in the water. The kind of transformation depends on the competition of influential factors: current depth and intensity of the bubble flux. There is a tendency to the transformation of methane bubbles in hydrate granular matter at low rate of bubbles flux and at relatively shallow depths in HSZ. There is a tendency to the formation of highly porous solid hydrate foam at intensive bubbles flux and at relatively deep depths. Perhaps future experimental observations shall reveal other forms of phase transformations for deep methane bubbles, and theoretical studies shall create the quantitative criteria for these forms.

# Acknowledgements

This work was supported by the Program of the Presidium of Russian Academy of Sciences № 23, the Subprogram "The nature of the World Ocean" of Federal Program "World Ocean" (Project 0013), the Fund for Protection of Lake Baikal and the RFBR grant 12-08-00067a. The authors thank V.E. Fortov for participation in deep-water experiments.

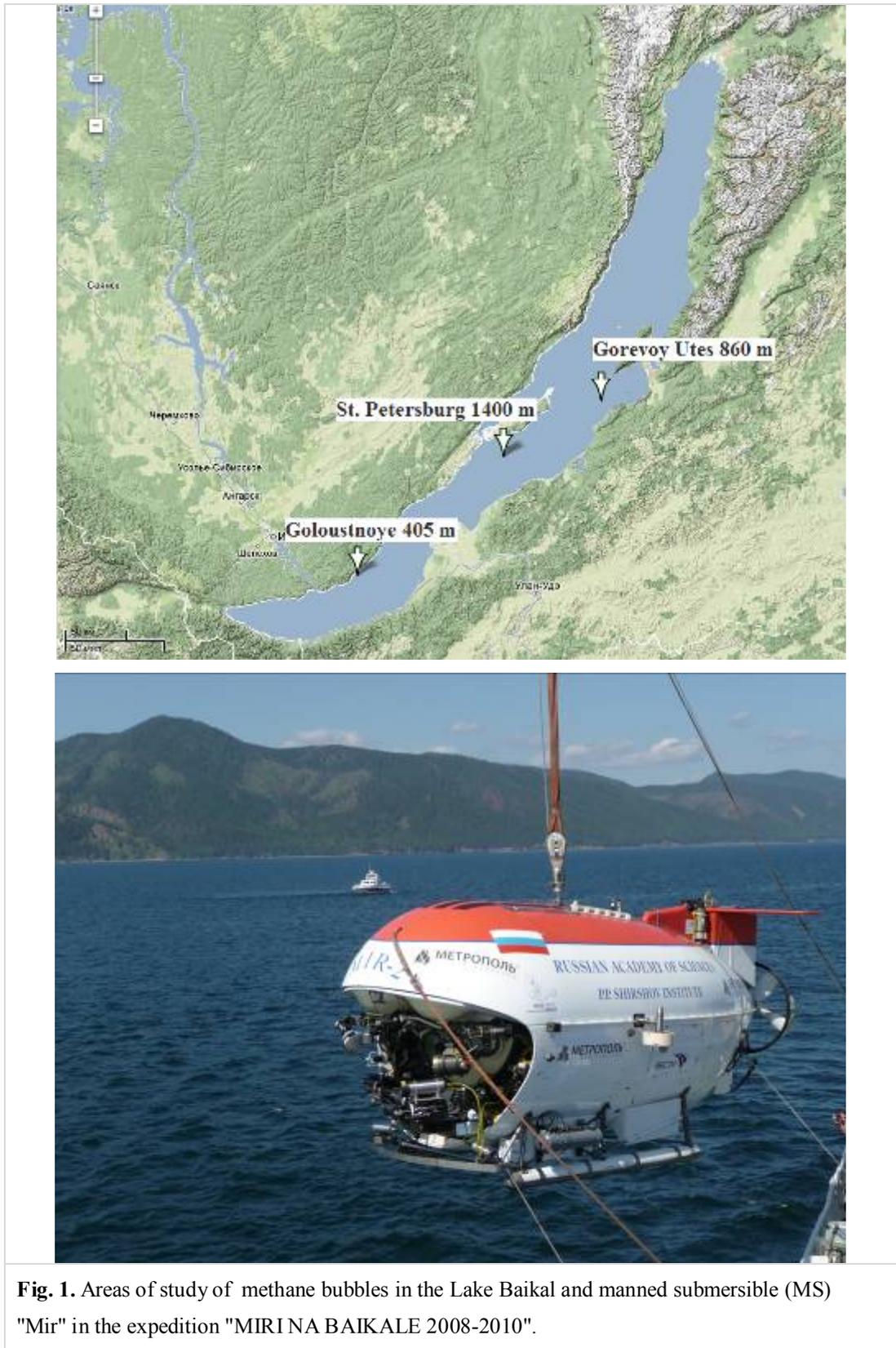

**Fig. 1.** Areas of study of methane bubbles in the Lake Baikal and manned submersible (MS) "Mir" in the expedition "MIRI NA BAIKALE 2008-2010".

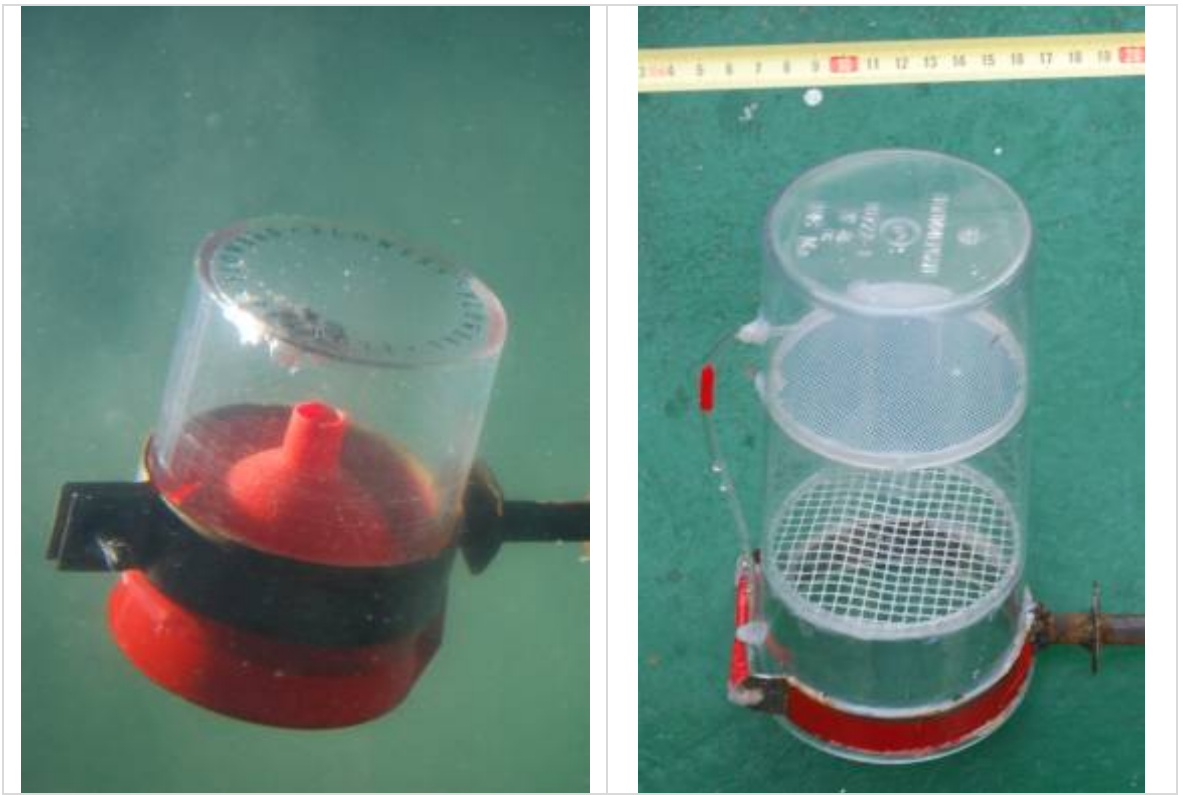

**Fig. 2.** Trap "Funnel" (85 mm diameter glass) and trap "Grid-2" (101 mm diameter glass, mesh sizes are 1 and 6 mm).

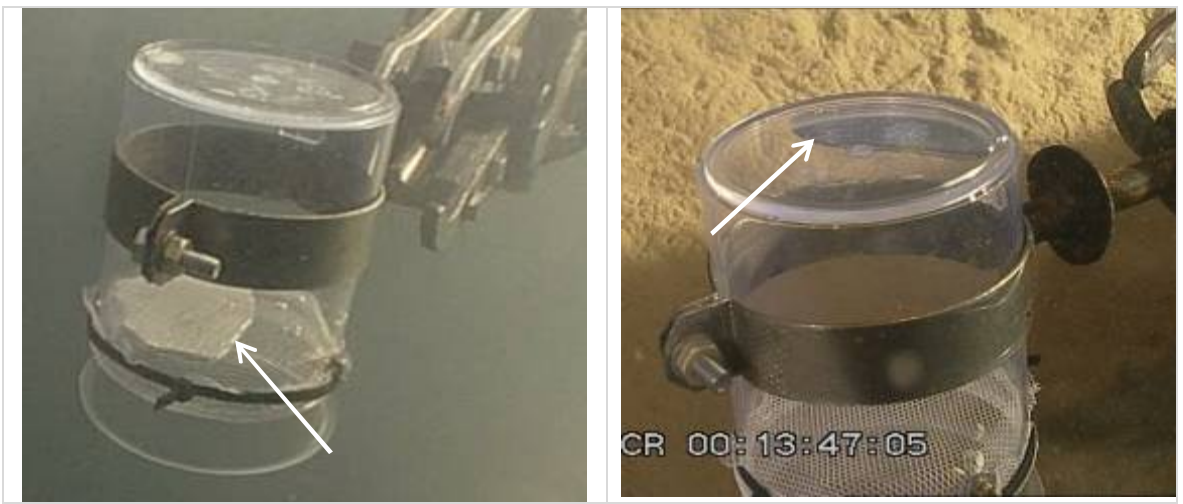

**Fig. 3.** Collection of gas in the trap "Grid-1" at a depth of 405 m. A single bubble (marked by arrow) is formed under the grid with a cell of 1 mm (left photograph). As a result of trap shaking the bubble flowed through the grid into the top section of the trap (right photograph).

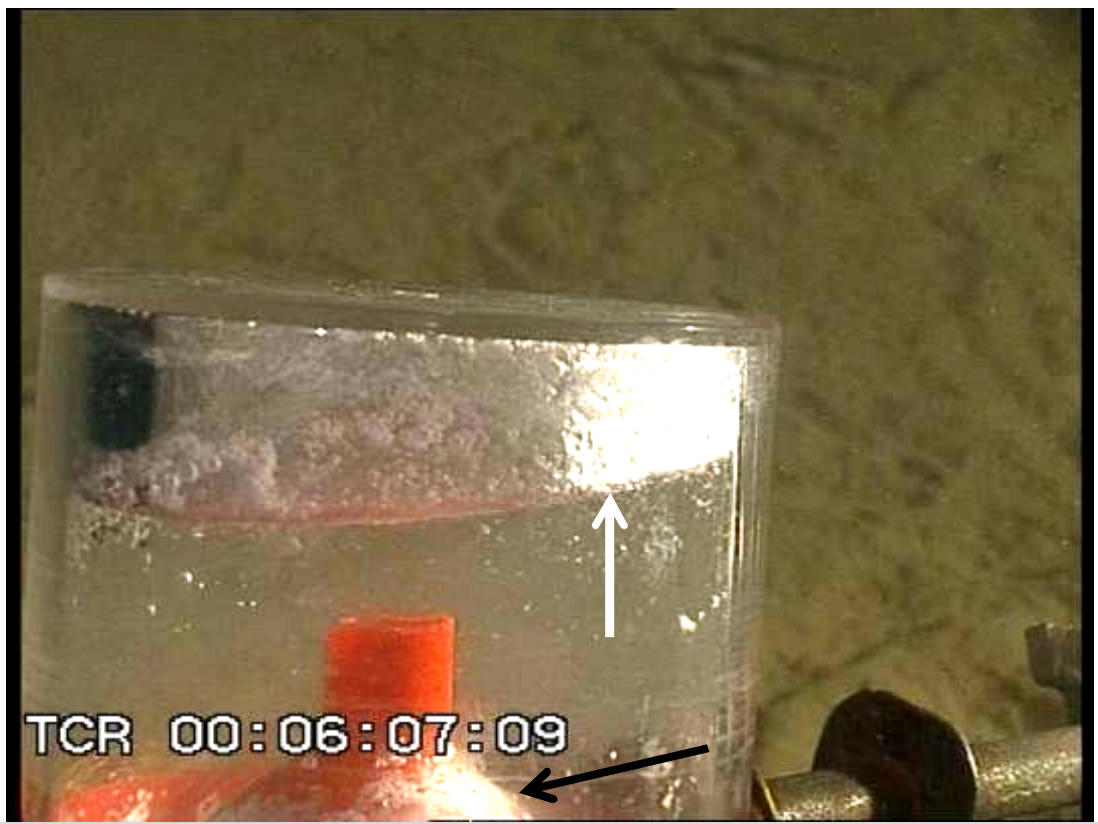

**Fig. 4.** Collection of gas by means of trap "Funnel". At a depth of 405 m collected gas accumulated in the upper part of the trap (meniscus is marked by white arrow). At a depth of 550 m wet inner surface (which was in gas phase) of the trap was covered by a layer of hydrate, as shown in the photo. At the earlier time the trap was overturned and hydrate formed on wet surface of the funnel (marked by black arrow).

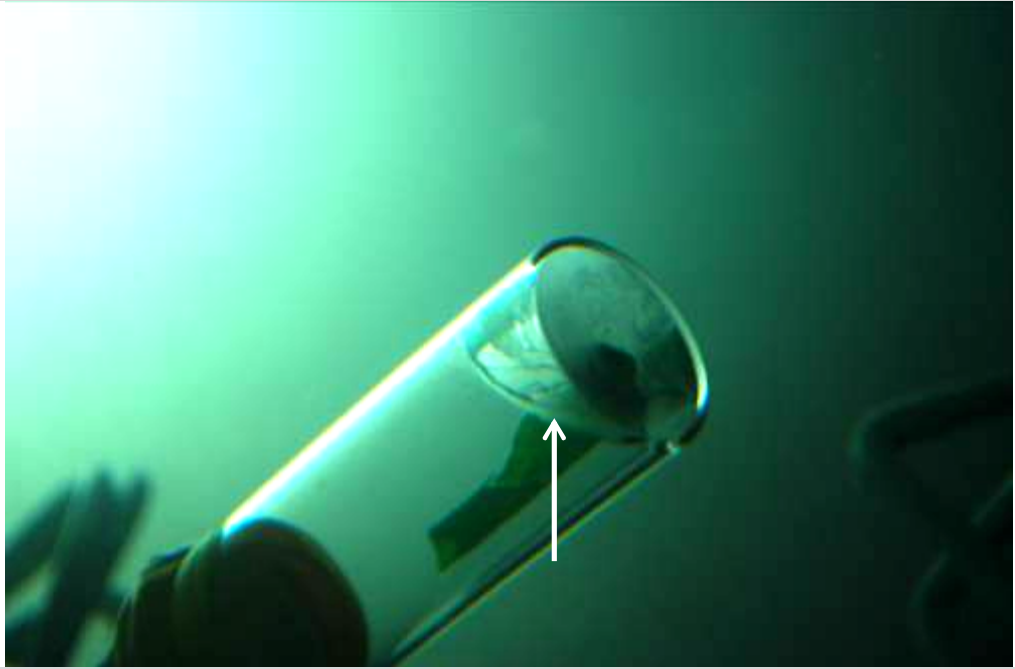

**Fig. 5.** Hydrate film (shown by arrow) was formed in the transparent bottle (~200 ml) with non-natural (industrial) methane at a depth of 1400 m.

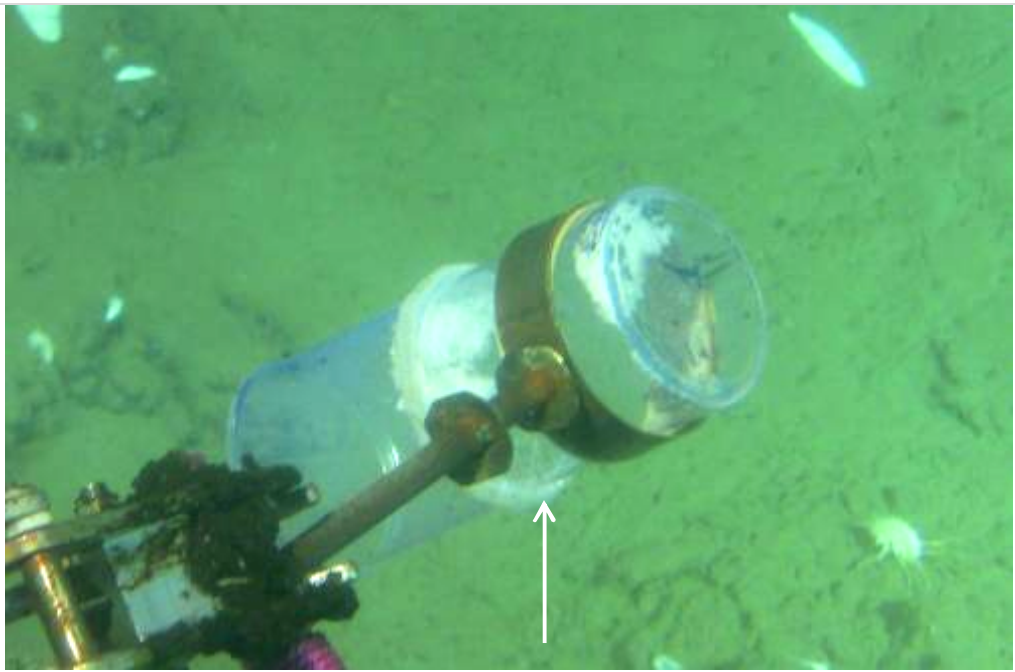

**Fig. 6.** Collection of methane bubbles with variant of trap "Grid-1" at a depth of 864 m. Grid with a mesh cell of 1 mm is mounted in the middle of cylindrical glass by means of white plastic ring (marked by arrow). Delayed by grid bubbles are collapsed and transformed into hydrate granular matter, part of which is sifted through a 1 mm mesh into the upper section of the trap.

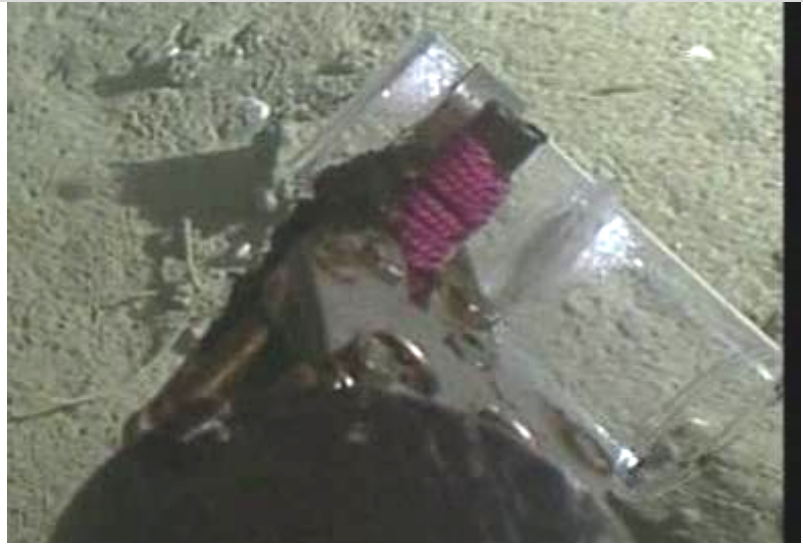

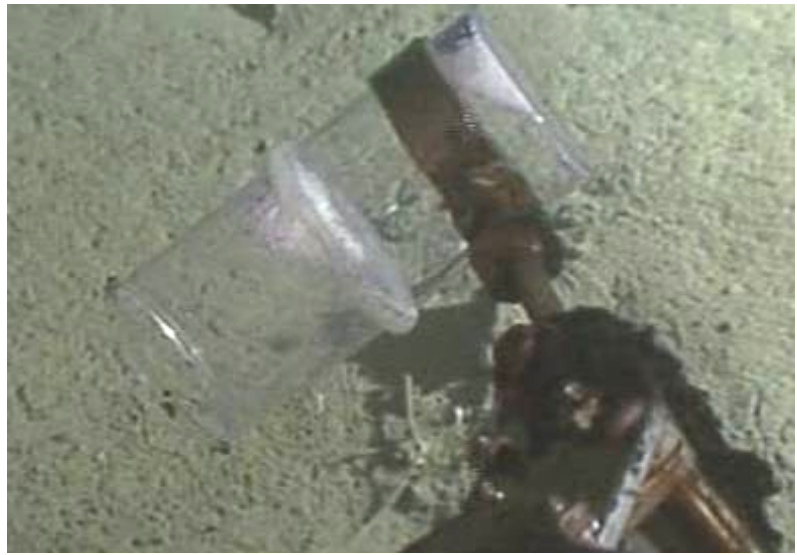

**Fig. 7.** Oscillations of the trap"Grid-1" at a depth of 864 m caused free interspersion of hydrate granular matter in the sections of the trap.

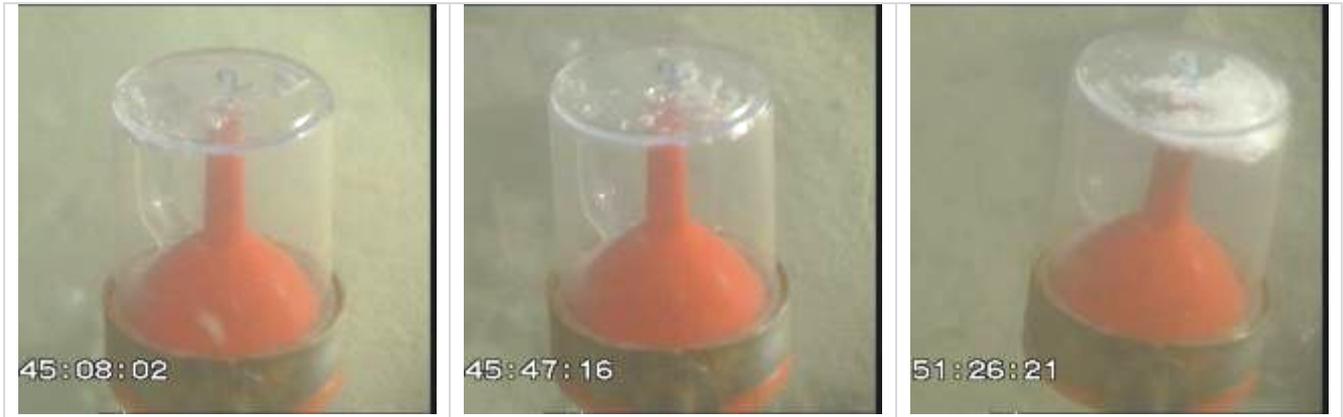

**Fig. 8.** Methane bubbles gradually transformed into granular matter in the trap "Funnel" at the depth of 864 m. The time-code on the Photographs is min: sec: sek100.

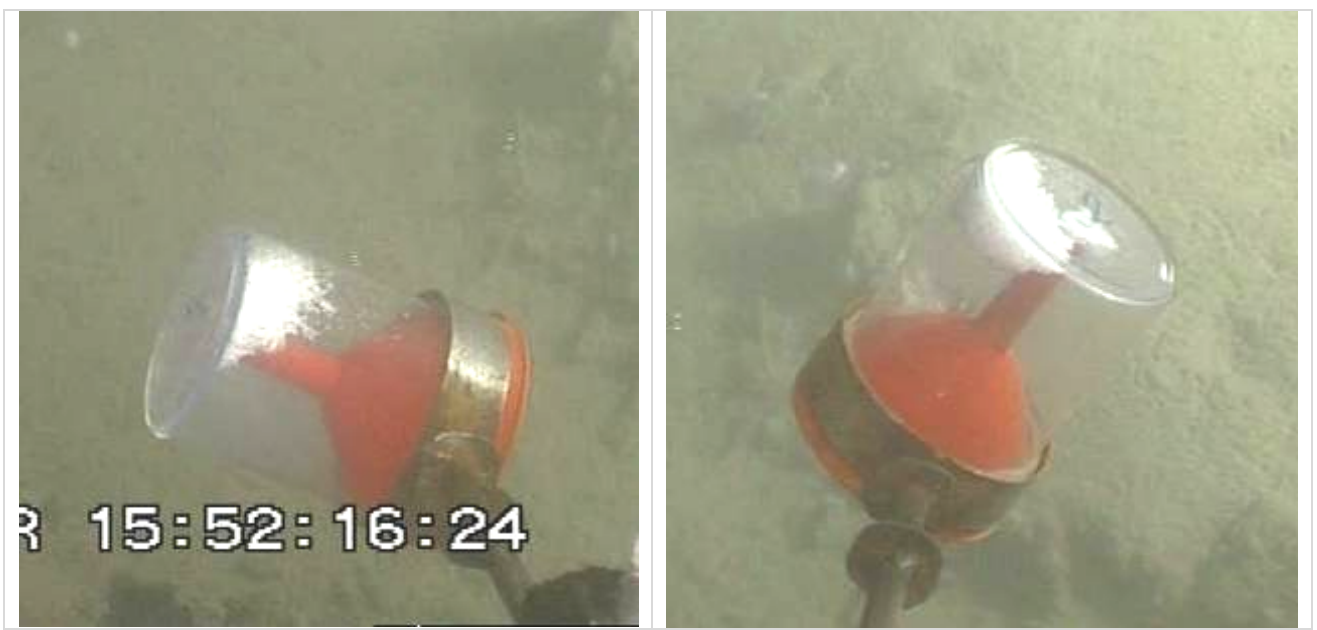

**Fig. 9.** The hydrate granular matter are interspersed freely from one end of trap to the another under oscillation of trap "Funnel" at the depth of 864 m.

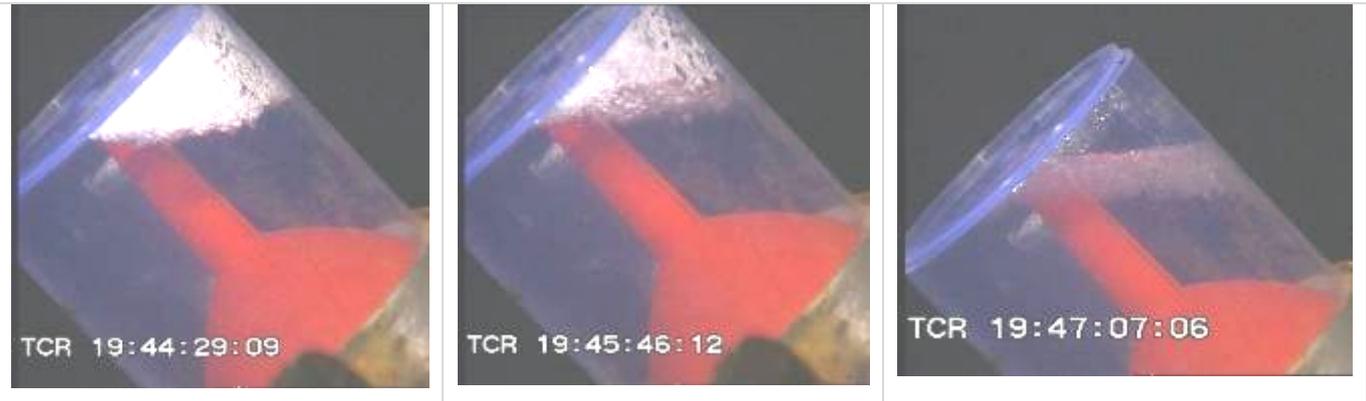

**Fig. 10.** The decomposition of hydrate granular material during a uniform MS lifting above HSZ. The local time is shown on photographs. At a depth of 387 m the first deformation of hydrate granular material in the trap is marked. At a depth of 325 m the almost complete transition into gas phase was observed. Snapshots present the trap content between these events. MS began surfacing from a depth of 860 m at 19:13 and MS reached the surface at 20:00.

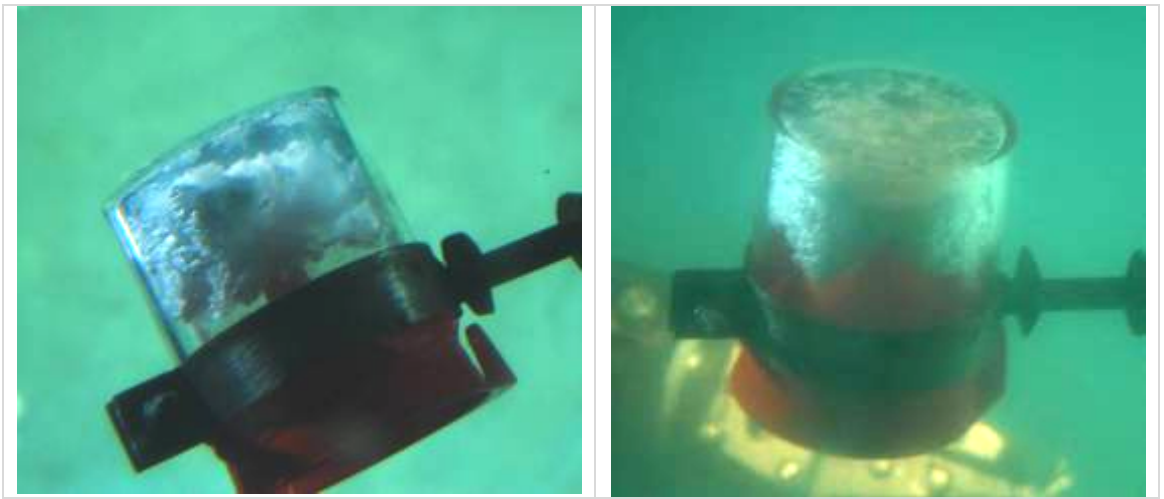

**Fig. 11.** Formation of solid methane hydrate foam in the "Funnel" at the depth of 1400 m. Photographs present the initial stage of filling of the trap (left) and half an hour later (right). Photographs were taken at 14:47 and 15:13 of the local time.

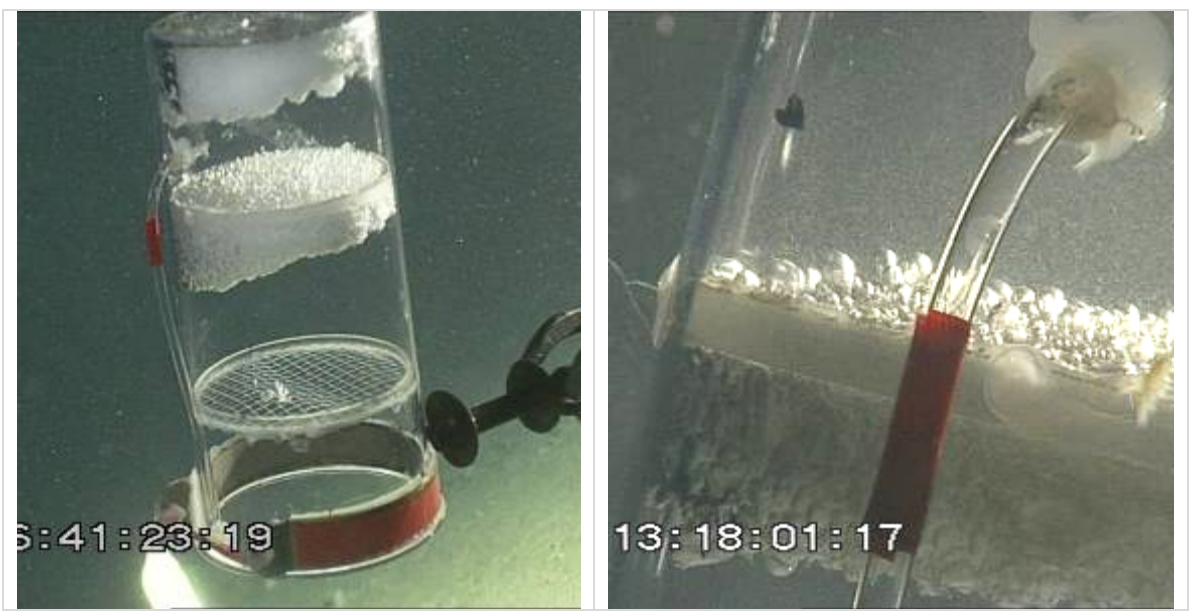

**Fig. 12.** Hydrate foam fills the trap "Grid-2" at a depth of 1400 m.

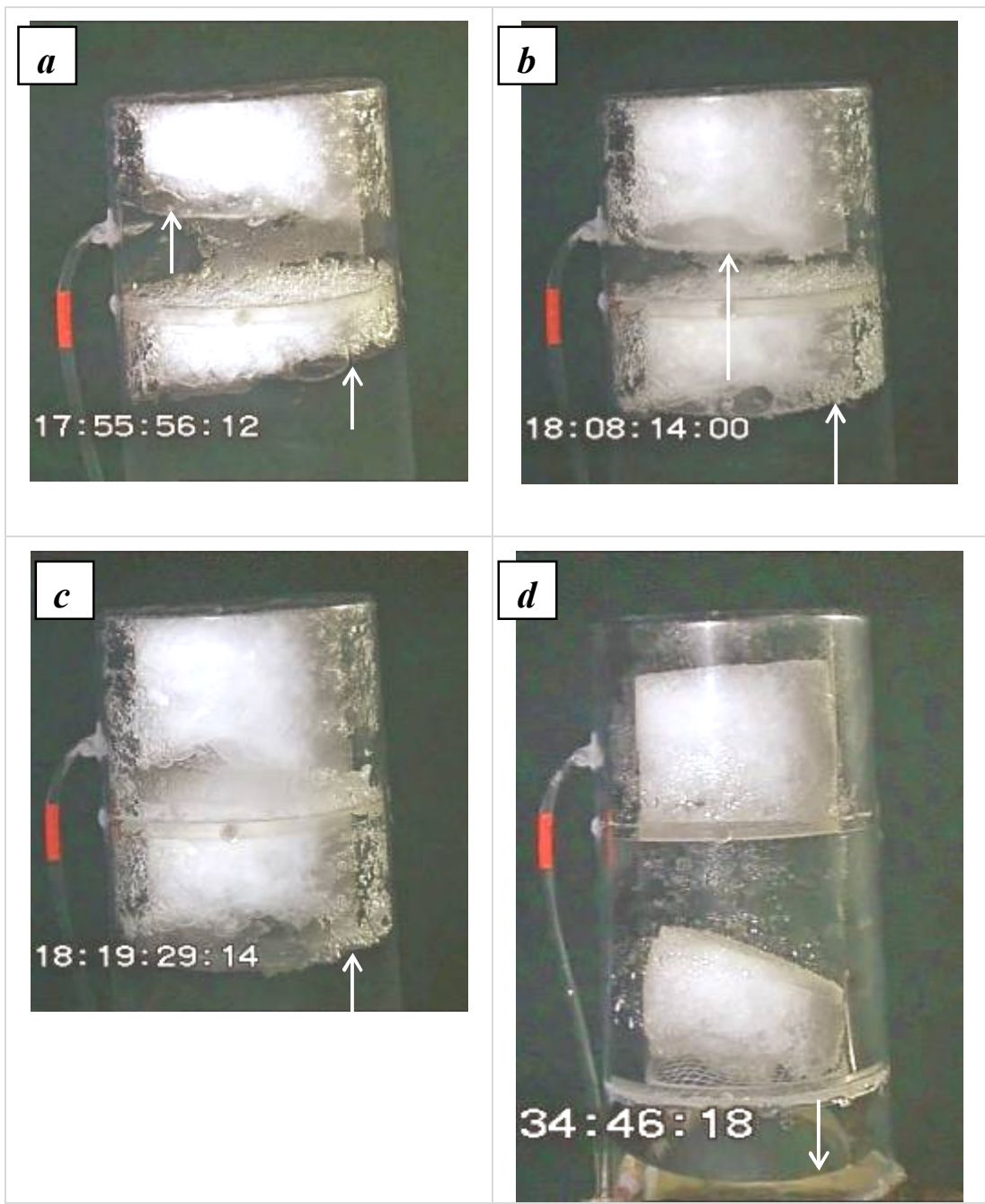

**Fig. 13.** During MS lifting the free gas is released from hydrate foams in upper and middle sections of the trap. Gas displaces water from the upper section as well as from the middle section of the trap. Pictures (*a-c*) show the upper and middle sections of the trap "Grid-2" during the lifting at a depth of 1029 m in 17:55 (*a*), at a depth of 776 m at 18:08 (*b*), at a depth of 561 m at 18:19 (*c*). At a depth of 245 m in 18:34 (*d*) all water is displaced by gas from the trap and pieces of solid hydrate foam fell on the lower grid with 6 mm cell. Arrows indicate the phase boundary between water and free gas.

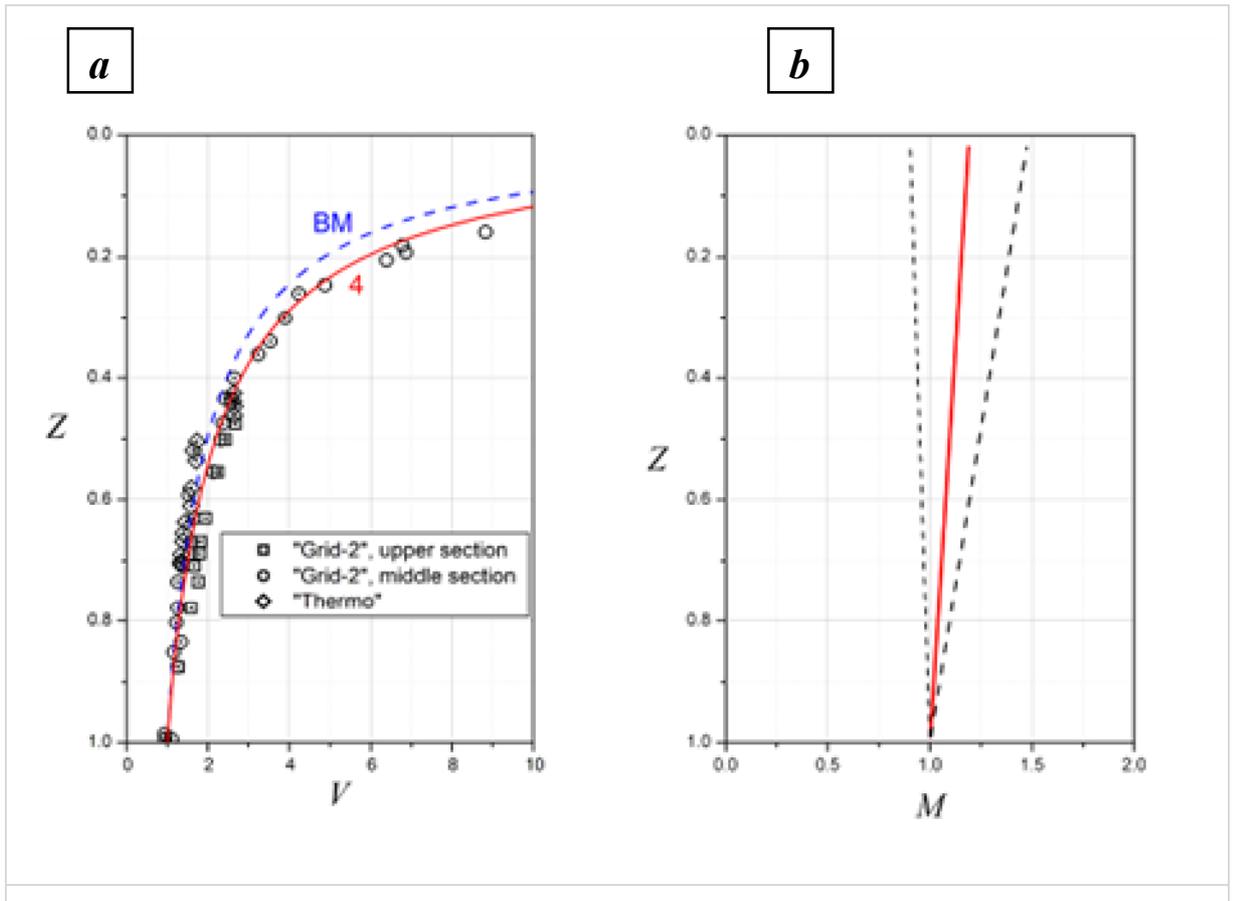

**Fig. 14.** (*a*) Variation of the dimensionless total volume of free gas and hydrate *V* in the traps "Thermo" and "Grid-2" during the lifting *V*=*V*(*Z*). Points within the symbols identify HSZ (*Z*<0.2714). Curve 4 is the approximation (4). Curve BM is the prediction of Boyle-Mariotte law $V=p_0/p$. (*b*) Variation of the dimensionless mass of the free gas *M* in the section of the trap during the lifting *M*=*M*(*Z*), calculated with the parameters (5). The dotted lines represent the confidence interval.

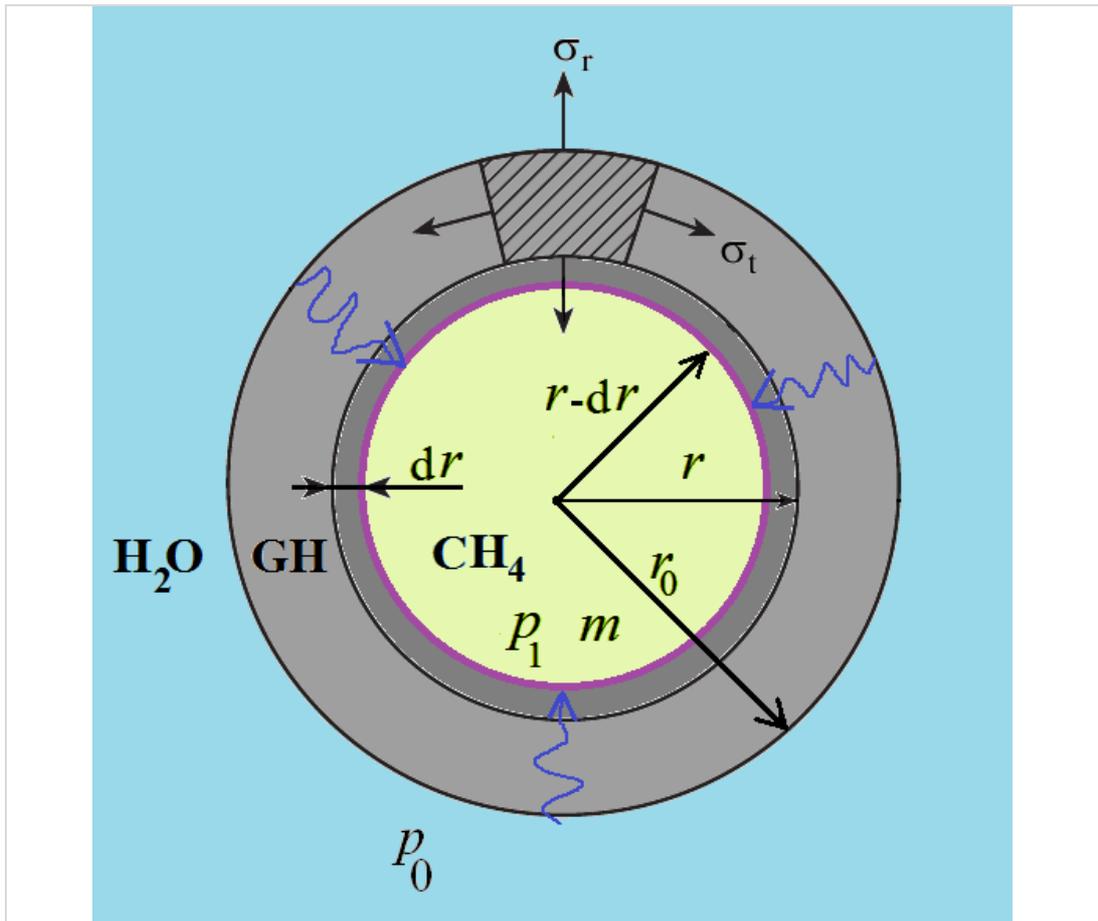

**Fig. 15.** The water from the ambient environment is filtered into the bubble, where the water comes into the reaction of hydration with methane. As a result the solid hydrate envelope is formed. Gas consumption in the bubble causes the pressure fall in the bubble and formation of the inner stresses in the envelope.

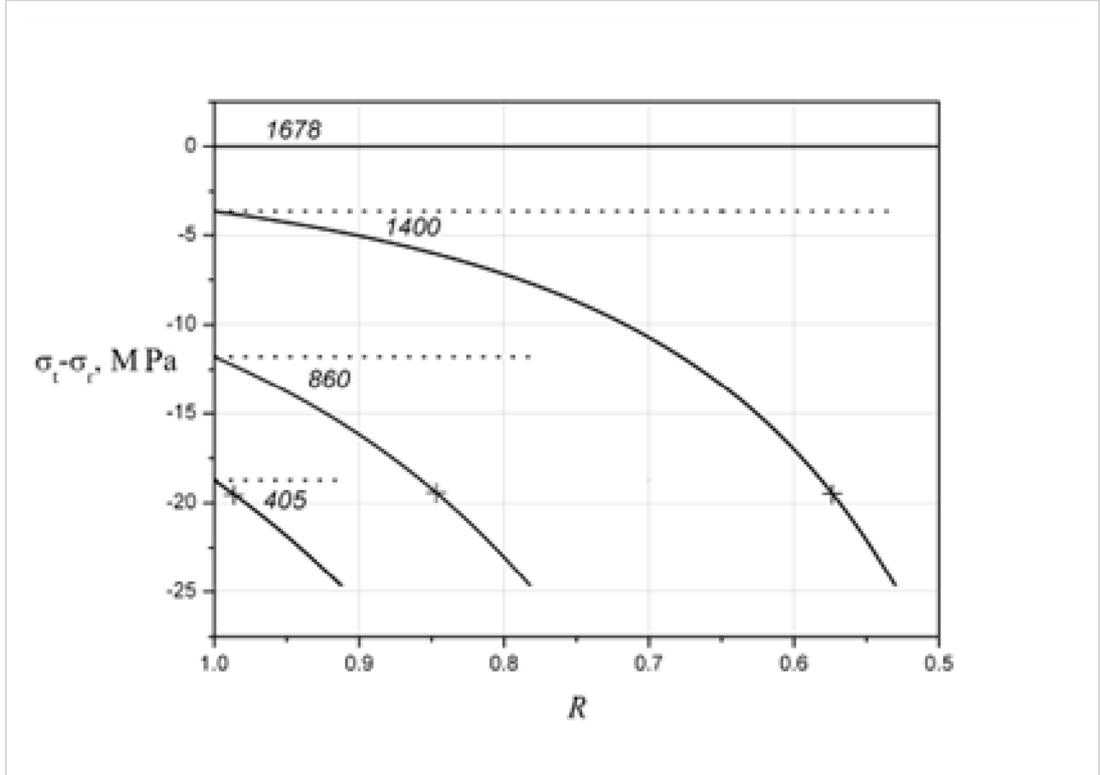

**Fig. 16.** Inner stresses in the hydrate envelope $\sigma_t-\sigma_r$ as a function of the envelope radius $R=r/r_0$ for depth $z_0$=405, 860, 1400 and 1678 m. Data for the internal surface of the envelope are shown by solid lines, data for external surface are shown by dashed ones. The lines show the stress until all gas in the bubble is transformed into hydrate (radius $R_f$). The crosses mark the appearance of thermodynamic instability of hydrate (radius $R_s$).

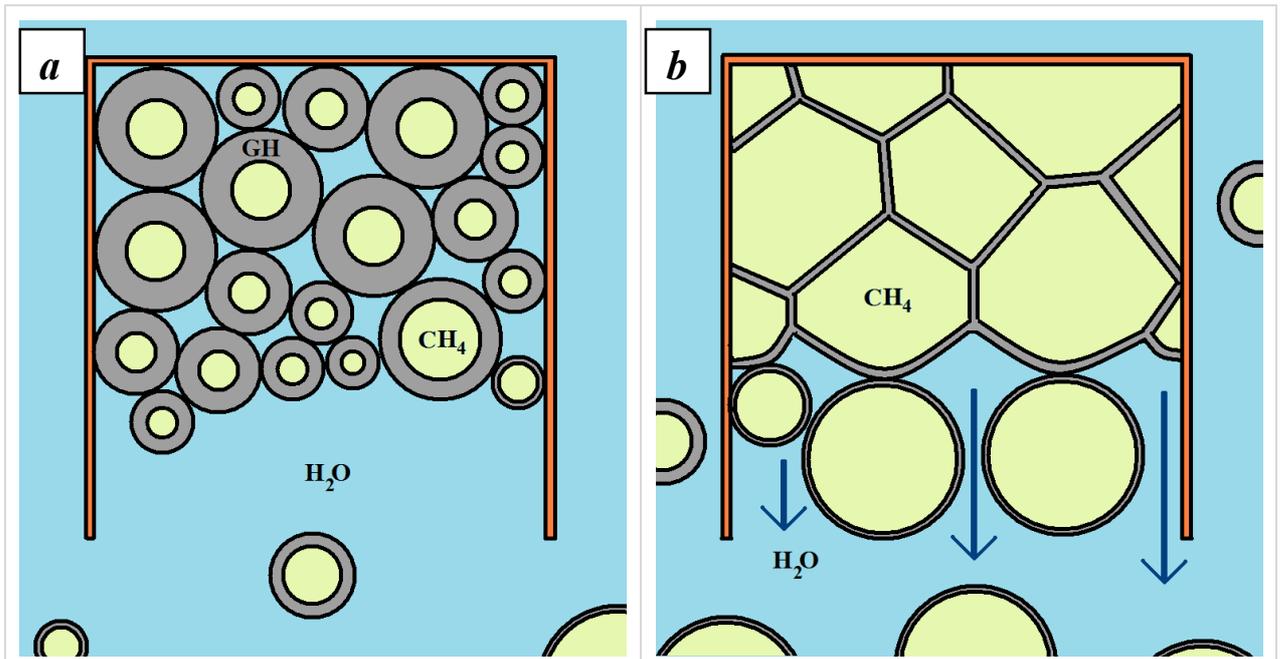

**Fig. 17.** Spherical solid hydrate foam (*a*) and polyhedral solid hydrate foam (*b*) in the trap.